\documentclass[aps,prb,showpacs,amsmath,twocolumn,amssymb]{revtex4-1}

\usepackage{graphicx}
\usepackage{dcolumn}
\usepackage{bm}
\usepackage{amsmath}
\usepackage{subfigure}
\usepackage{color}
\usepackage{bbding}
\usepackage{pifont}

\begin{document}


\title{Superconducting proximity effect in semiconductor thin films with spin-splitting and spin-orbit interaction}%

\author{J. Michelsen}

\author{R. Grein}
\affiliation{%
Intitut f\"{u}r Theoretische Festk\"{o}rperphysik, Karlsruhe Institute of Technology, D-76131 Karlsruhe, Germany
}%



\begin{abstract}
Superconducting heterostructures with spin-active materials have emerged as promising platforms for engineering topological superconductors featuring Majorana bound states at surfaces, edges and vortices. 
Here we present a method for evaluating, from a microscopic model, the band structure of a semiconductor film of finite thickness deposited on top of a conventional superconductor. Analytical expressions for the proximity induced gap openings are presented in terms of microscopic parameters and the proximity effect in presence of spin-orbit and exchange splitting is visualized in terms of Andreev reflection processes. An expression for the topological invariant, associated with the existence of Majorana bound states, is shown to depend only on parameters of the semiconductor film. The finite thickness of the film leads to resonant states in the film giving rise to a complex band structure with the topological phase alternating between trivial and non-trivial as the parameters of the film are tuned.
\end{abstract}

\maketitle
%
%
%

A topological superconductor is a superconductor with a non-trivial topological order and associated zero energy edge-, surface- or defect modes. Recently\cite{Schnyder2008}, a classification of the possible distinct topological phases of superconductors (and insulators), characterized by an integer topological invariant, was constructed following an earlier classification scheme based on the symmetric spaces of Bogoluibov-de Gennes (BdG) Hamiltonians due to Altland and Zirnbauer\cite{Altland1997}. One of the most interesting properties of the more exotic classes of superconductors is that the creation operators associated with the zero energy boundary modes are the same as their annihilation operators, leading to topologically degenerate ground states. Such modes, commonly referred to as Majorana fermions, constitute essential ingredients in proposals of schemes for fault tolerant quantum computing known as topological quantum computing\cite{Nayak2008}. The prototypical example of such a system\cite{Ivanov2001}, the $p_x+ip_y$ superconductors, may be \textit{intrinsically} realized in the 2D layered material Sr$_2$RuO$_4$. However, since time-reversal symmetry is spontaneously broken, the $p_x+ip_y$ and $p_x-ip_y$ phases are degenerate which generally leads to formation of domains with both chiralities, thus complicating the edge-state structure. Furthermore, the layered structure, as well as the low intrinsic gap at higher temperatures implies that Majorana bound states at e.g. vortices are only protected by a very small minigap in the mK range. These considerations contributed to the general excitement following several proposals for \textit{engineered} topological superconductors, where heterostructures involving conventional s-wave superconductors and various spin-active materials are used to produce artificial lower dimensional topological superconductors\cite{Fu2009,Alicea2010,Brouwer2011,Alicea2011,Beenakker2012,Tanaka2012}. 
While the topological invariant of the "intrinsic" $p_x\pm ip_y$ superconductors can be evaluated directly from the Bloch-functions (or equivalently from the Bloch-Hamiltonian)\cite{Schnyder2008} of the periodic structure, the "engineered" topological superconductors are highly non-homogenous in at least one spatial direction which complicates a direct evaluation of the topological invariant. The situation is simplified by phenomenologically modeling the system by adding an effective "pair-potential" to the uncoupled Hamiltonian of the non-superconducting material to account for the proximity effect which opens a gap due to the mixing of electrons and holes in the Andreev reflection process. Indeed such a model can be justified\cite{Sau2010,Alicea2012} by an approach originally due to McMillan\cite{McMillan1968}, where the hybrid structure is described by a tunnel model, and elimination of the superconductor leads to an effective Hamiltonian of the system with an effective "pair-potential". While providing a very elegant qualitative derivation of the model, it should be noted that this is still a somewhat crude (depending on the detailed description of the tunneling terms) model of the proximity effect which obscures the role of Andreev reflection and may miss crucial aspects of the gap dependence on thickness, barrier height etc.

In this article we develop a framework to examine the spectral properties of proximity induced surface states in thin films on conventional superconductors. The film is assumed to be non-superconducting but generally spin-active. Aiming to retain as much generality as possible, the thin film is initially treated as a "black box", characterized by a "reflection matrix" $\boldsymbol{R}$, which acts as a matching condition for quasiparticle wave functions decaying into the superconductor (Andreev bound states/deGennes-Saint James states). Solvability of the matching conditions provide a spectral equation which allows us to obtain the effective 2D bandstructure $\varepsilon_n(\boldsymbol{k}_{||})$ (translational invariance in the directions parallel to the film is assumed).The reflection matrix is a $4\times 4$ unitary matrix with diagonal entries in electron-hole space. 
\begin{equation}\label{reflectionmatrix}
\boldsymbol{\Phi}^\text{out}=\boldsymbol{R}\boldsymbol{\Phi}^\text{in}, \quad \boldsymbol{R}(\boldsymbol{k}_{||},\varepsilon)=\begin{pmatrix}
\boldsymbol{r}_e(\boldsymbol{k}_{||},\varepsilon) & \boldsymbol{0}\\
\boldsymbol{0} & \boldsymbol{r}_h(\boldsymbol{k}_{||},\varepsilon)\end{pmatrix}.
\end{equation}
The matrices are subject to symmetry constraints, in particular the inherent particle-hole symmetry originating from a redundant description of the system
\begin{equation}\label{particlehole}
\boldsymbol{r}_h(\boldsymbol{k}_{||},\varepsilon)=\boldsymbol{r}_e^T(-\boldsymbol{k}_{||},-\varepsilon),
\end{equation}
as well as possible "true" symmetries such as time-reversal symmetry,
\begin{equation}\label{timerev}
\boldsymbol{r}_e(\boldsymbol{k}_{||},\varepsilon)=\sigma_y \boldsymbol{r}^T_e(-\boldsymbol{k}_{||},\varepsilon)\sigma_y,
\end{equation}
and spin-rotational symmetry
\begin{equation}\label{spinrot}
\boldsymbol{r}_e(\boldsymbol{k}_{||},\varepsilon)=r_e(\boldsymbol{k}_{||},\varepsilon)\boldsymbol{1},
\end{equation}
with $|r_e|^2=1$ due to unitarity.
As will be shown in the next section, solvability of the matching condition of the exponentially decaying solutions on the superconducting side leads to an expression:
\begin{equation}
0=\text{det}\left(\boldsymbol{1}-\boldsymbol{r}_e^{-1}\sigma_y e^{-i\gamma}\boldsymbol{r}_h \sigma_y e^{-i\gamma}\right),
\end{equation}
with $\gamma=\arccos(\varepsilon/\Delta)$ the Andreev reflection phase shift. The occurrence of the $\sigma_y$ matrices reflects the nature of the Andreev reflection mechanism in a singlet superconductor ($e\uparrow \rightarrow h\downarrow$ and $e\downarrow \rightarrow -h\uparrow$). The solutions of the spectral equation yields the 2D bandstructure $\varepsilon_n(\boldsymbol{k}_{||})$. We derive a representation of the spectral equation that allows for an intuitive interpretation of the spectrum in terms of quantization rules akin to the semiclassical Born quantization rules. It should be emphasized, however, that this calculation is exact at this level. This representation shall be used extensively in section \ref{classes} where the reflection matrices are classified according to the scheme developed by Altland \cite{Altland1997}, and others \cite{Schnyder2008,Ryu2010}. There we shall also make general observations of the various symmetry classes and consider some representative examples which we analyze numerically as well as analytically to some detail.

The central objective of this work is to find a microscopic description of the class of hetero-structures involving spin-orbit coupled semiconductors with additional spin-splitting, originating either from an applied magnetic field or proximity induced exchange coupling, and conventional superconductors believed to have topologically non-trivial bandstructure for a certain regime of parameters, and associated Majorana bound states at the edges of the system \cite{Alicea2010,Brouwer2011,Alicea2011}. For this class of systems, the aforementioned spectral equation reads:
\begin{equation*}\begin{split}
0=&\cos^2\theta\sin(\gamma-\delta\raisebox{1pt}{$\chi$}^s_{+})\sin(\gamma-\delta\chi^s_{-})\\
+&\sin^2\theta\sin(\gamma-\delta\chi^t_{+})\sin(\gamma-\delta\chi^t_{-}),
\end{split}
\end{equation*}
where $\gamma=\arccos(E/\Delta)$ is the phase shift from the Andreev reflection process, $2\delta\chi^s_{\pm}=\chi_{e\pm}-\chi_{h\mp}$ represent the differences in scattering phase shifts for electron-hole branches off opposite spin/helicity (in this notation $s$ is for singlet-like), while $2\delta\chi^t_{\pm}=\chi_{e\pm}-\chi_{h\pm}$ represent the differences in scattering phase shifts for electron-hole branches of the same helicity branch ($t$ for equal-spin triplet-like). The rotation angle, $\theta$, parametrizing the diagonalizing matrix, $\boldsymbol{U}$, describes the relative magnitude of spin-orbit coupling vs exchange splitting ($\theta=0$ in the absence of spin-orbit coupling and $\theta=\pi/2$ in the absence of exchange splitting).

The analysis of the spectral equation reveals that the system is always fully gapped for $\boldsymbol{k}_{||}\neq 0$ as long as spin-orbit interaction is present. At $\boldsymbol{k}_{||}=0$ (where spin-orbit interaction is zero in any case), the gap opens and closes upon tuning the exchange splitting. The opening and closing of the gap signals a topological phase transition, in the sense that two distinct phases (bandstructures) can not be adiabatically (smoothly) connected without closing the gap (at critical point in parameter space corresponding to the topological phase transition). The topological invariant characterizing the different phases is shown to be
\begin{equation*}
\mathcal{M}=\text{sign}\left[\cos\delta\chi^0\right]\in \{-1,1\}\simeq \mathbb{Z}_2,
\end{equation*}
where $\delta\chi^0\equiv\delta\chi^s_{+}(\varepsilon=0,\boldsymbol{k}_{||}=0)=-\delta\chi^s_{-}(\varepsilon=0,\boldsymbol{k}_{||}=0)$. Physically, a topological phase transition occurs whenever one of the branches is pushed below zero energy which thus changes the number (more specifically the parity) of occupied bands. This interpretation is consistent with the recent work by Potter and Lee \cite{Potter2011a}, as well as Stanescu, DasSarma et. al.\cite{Stanescu2011}, where they study the influence of multiple bands (due to quantized transverse modes) starting from the phenomenological model of the system. In our description multiple bands are an intrinsic feature of the microscopic model and the finite thickness of the film. 

\begin{figure}
\includegraphics[width=0.45\textwidth]{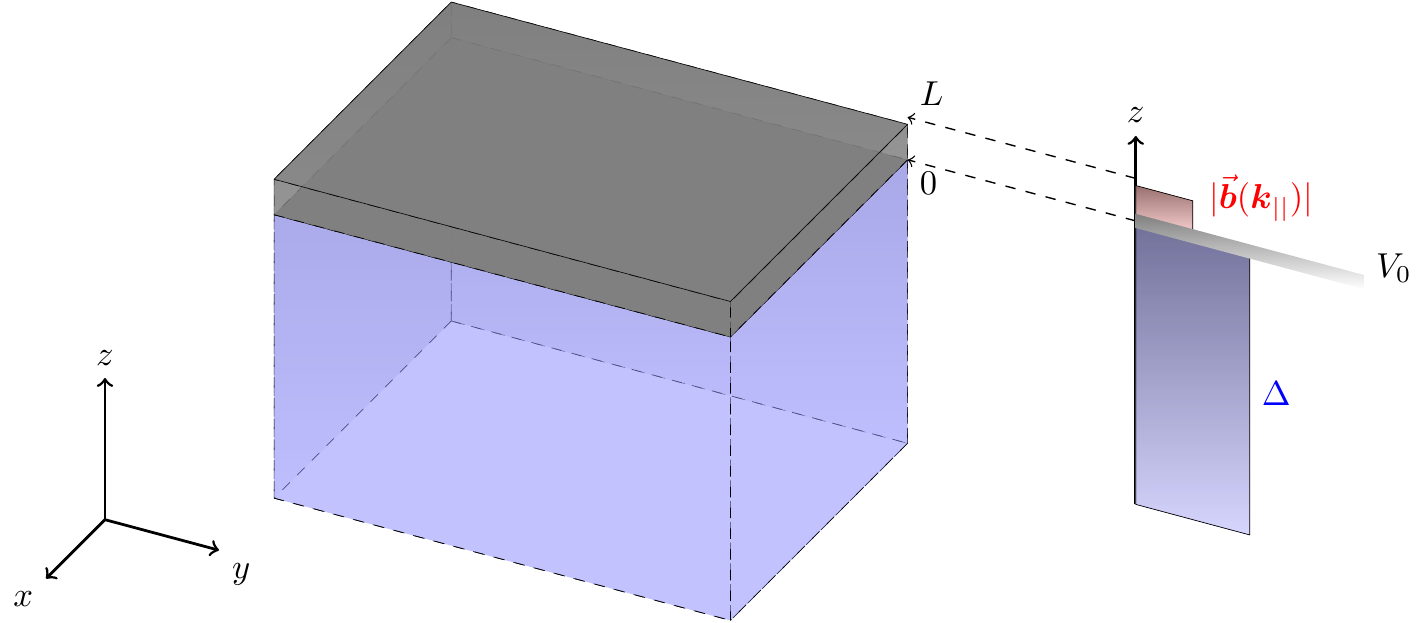}
\caption{Schematic illustration of the system considered in this article: A bulk superconductor with a film of finite thickness, $L$, deposited on top. Translational invariance parallel to the film is assumed, while the Hamiltonian in the superconducting region differers from that in the film region in that the superconducting order parameter $\Delta$ is nonzero only in the superconductor, while spin-dependent terms, here represented by a term $\boldsymbol{b}(\boldsymbol{k}_{||})\cdot\boldsymbol{\sigma}$ in the normal state Hamiltonian, are nonzero only in the film region. An interface potential, in the following modeled as a delta-function with weight $V_0$, is also assumed to be present.}
\end{figure}


\section{Spectral Equation}
The type of system in mind is a conventional (s-wave, singlet) superconductor with a non-superconducting film of arbitrary thickness deposited on top of it. Anticipating the film to break spin-rotational symmetry we extend the usual two-component Nambu spinor to its four-component form
\begin{equation}
\boldsymbol{\Psi}=\begin{pmatrix}
\boldsymbol{\psi}\\
\boldsymbol{\psi}^*\end{pmatrix}
=\begin{pmatrix}
\psi_\uparrow\\
\psi_\downarrow\\
\psi^\dag_\uparrow\\
\psi^\dag_\downarrow\end{pmatrix}
\end{equation}
where, by slight abuse of notation, we denote $\psi^*=(\psi^T)^\dag$.
Due to the redundancy introduced in this notation, the four-component spinor is pseudo-real in the sense that $\Psi=\tau_x\Psi^*$. This structure of the spinor implies a constraint on the Hamiltonian
\begin{equation}
\boldsymbol{H}=-\tau_x\boldsymbol{H}^*\tau_x,
\end{equation}
usually referred to as particle-hole-symmetry. By this constraint the Hamiltonian then has the general structure
\begin{equation}
\boldsymbol{H}=\begin{pmatrix}
\boldsymbol{\mathcal{H}} & \boldsymbol{\Delta}\\
-\boldsymbol{\Delta}^* & -\boldsymbol{\mathcal{H}}^*
\end{pmatrix},
\end{equation}
where $\boldsymbol{\mathcal{H}}$ is the normal state Hamiltonian and $\Delta_{\sigma\sigma'}=-g \langle\psi_\sigma\psi_{\sigma'}\rangle$ represents the superconducting pair potential that should, in principle, be evaluated self-consistently. The attractive interaction $-g$ is non-zero only in the superconducting region, implying that the pair-potential is zero in the film region, while the assumption of a singlet superconductor fixes the spin-structure of the pair-potential in the superconducting region $\Delta_{\sigma\sigma'}=(i\sigma_y)_{\sigma\sigma'}\Delta$. Together with the translational invariance along the film, the Hamiltonian in the superconducting region takes the form
\begin{equation}
\boldsymbol{H}(z<0)=\begin{pmatrix}
\xi_{||}(-i\partial_z) & i\sigma_y\Delta\\
-i\sigma_y\Delta & -\xi_{||}(-i\partial_z)\end{pmatrix}.
\end{equation}
where $\xi_{||}(-i\partial_z)=\tfrac{(-i\partial_z)^2}{2m}-\mu_{||}$ with $\mu_{||}=\mu-\frac{\boldsymbol{k}_{||}^2}{2m}$. The elementary solutions of this Hamiltonian are plane waves, $\boldsymbol{\Phi}(\boldsymbol{k}_{||},k)e^{ikz}$ with the envelopes satisfying the eigenvalue equation
\begin{equation}\label{BdGequation}
\boldsymbol{\Phi}(\boldsymbol{k}_{||},k)=\begin{pmatrix}
\boldsymbol{u}\\
\boldsymbol{v}\end{pmatrix}: \quad 
\begin{matrix}
\xi_{||}(k)\boldsymbol{u}+\Delta i\sigma_y\boldsymbol{v}=\varepsilon\boldsymbol{u}\\
\end{matrix},
\end{equation}
with a wave vector determined from the condition
\begin{equation}\label{BdGdispersion}
\varepsilon^2=\xi^2_{||}(k)+\Delta^2.
\end{equation}
Since the right hand side is a fourth order polynomial in $k$ we generally have 4 different solutions. 
For $\varepsilon^2<\Delta^2$ we must have $\text{Re}(\xi_{||}(k))=0$. By direct inspection we see that if $k$ is a solution, then so are $-k$, $+k^*$ and $-k^*$. The wave functions corresponding to these wave vectors are either exponentially growing or decaying into the bulk of the superconductor. For a semi-infinite geometry only the exponentially decaying solutions are physical, and we must have $\text{Im}k<0$. Denoting the solution with positive real part simply by $k_+$ we note that $k_-=-k_+^*$ is also a physical solution, while the other two are non-physical (not normalizable). 
Let us also introduce the phase $\gamma=\arccos(\varepsilon/\Delta)$ such that $\cos\gamma=\varepsilon/\Delta$ and $\sin\gamma=|\xi_{||}(k)|/\Delta=\sqrt{1-(\varepsilon/\Delta)^2}$. 
With this parametrization Eq. (\ref{BdGequation}) obtains the form
\begin{equation}
\boldsymbol{v}^\pm=-i\sigma_ye^{\mp i\gamma}\boldsymbol{u}^\pm,
\end{equation}
where $\boldsymbol{u}^{\pm}=\boldsymbol{u}(k_\pm)$ and $\boldsymbol{v}^{\pm}=\boldsymbol{v}(k_\pm)$. The full wave-function on the superconducting side is a linear combination
\begin{equation}
\boldsymbol{\Phi}(\boldsymbol{k}_{||},z)=\boldsymbol{\Phi}^{\text{in}}(\boldsymbol{k}_{||},z)\frac{e^{ik_+z}}{\sqrt{v}}+\boldsymbol{\Phi}^\text{out}(\boldsymbol{k}_{||},z)\frac{e^{ik_- z}}{\sqrt{v}},
\end{equation}
where $v=\text{Re}(k_\pm)/m$ and we have associated the wave function with positive real part of the wave vector as an "incoming" wave and the wave function with negative real part of the wave vector as "outgoing" wave. Specifically we have
\begin{equation}
\boldsymbol{\Phi}^\text{in}=\begin{pmatrix}
\boldsymbol{u}^+\\
-i\sigma_ye^{-i\gamma}\boldsymbol{u}^+\end{pmatrix}, \qquad \boldsymbol{\Phi}^\text{out}=
\begin{pmatrix}
\boldsymbol{u}^-\\
-i\sigma_ye^{i\gamma}\boldsymbol{u}^-\end{pmatrix}.
\end{equation}
Applying the boundary condition $\boldsymbol{\Phi}^\text{out}=\boldsymbol{R}\boldsymbol{\Phi}^\text{in}$ between the incoming and outgoing wave functions through the scattering matrix $\boldsymbol{R}$ from Eq. (\ref{reflectionmatrix}) we get
\begin{equation}
\begin{split}
\boldsymbol{u}^-&=\boldsymbol{r}_e\boldsymbol{u}^+,\\
(i\sigma_y) e^{i\gamma}\boldsymbol{u}^-&=\boldsymbol{r}_h(i\sigma_y) e^{-i\gamma}\boldsymbol{u}^+.
\end{split}
\end{equation}
Combining the two finally leads to the equation
\begin{equation}
0=(1-\boldsymbol{r}_e^{-1}\sigma_ye^{-i\gamma}\boldsymbol{r}_h\sigma_ye^{-i\gamma})\boldsymbol{u}^+.
\end{equation}
The solvability of this equation then implies the spectral equation
\begin{equation}
0=\text{det}\Bigl(\boldsymbol{1}-\boldsymbol{r}^{-1}_e(\boldsymbol{k}_{||},\varepsilon)\sigma_ye^{-i\gamma(\varepsilon)}\boldsymbol{r}_h(\boldsymbol{k}_{||},\varepsilon)\sigma_ye^{-i\gamma(\varepsilon)}\Bigr).
\end{equation}
The solutions of this equation constitute the bound state spectrum, or 2D bandstructure $\varepsilon_n(\boldsymbol{k}_{||})$, of the states localized around the film region.

It is useful to point out that a gapped system can be identified by the condition (note $\gamma(0)=\pi/2$):
\begin{equation}\label{isitgapped}
0\neq \text{det}\Bigl(\boldsymbol{1}+\boldsymbol{r}_e^{-1}(\boldsymbol{k}_{||},0)\sigma_y\boldsymbol{r}_h(\boldsymbol{k}_{||},0)\sigma_y\Bigr), \quad \forall \boldsymbol{k}_{||}.
\end{equation}

\section{Symmetry Classification}\label{classes}
In this section we shall discuss the properties of the reflection matrix $\boldsymbol{r}_e$. Note that by virtue of the inherent redundancy, here referred to as particle-hole symmetry, the properties of $\boldsymbol{r}_e$  directly determine those of $\boldsymbol{r}_h$. In order to systematically analyze the properties we follow the classification scheme of Altland and Zirnbauer.
\begin{center}
\begin{tabular}{|c|c|c|}
\hline
\multicolumn{3}{|c|}{\textbf{Classification of Superconductors}}\\ \hline
\textbf{Class} & Time-rev. & Spin-rot.  \\ \hline \hline
D & \textcolor{red}{\ding{56}} & \textcolor{red}{\ding{56}} \\ \hline
C & \textcolor{red}{\ding{56}} & \textcolor{green}{\ding{52}} \\ \hline
DIII & \textcolor{green}{\ding{52}} & \textcolor{red}{\ding{56}} \\ \hline
CI & \textcolor{green}{\ding{52}} & \textcolor{green}{\ding{52}} \\ \hline
\end{tabular}
\end{center} 
Let us first discuss the effect of time-reversal symmetry, Eq. (\ref{timerev})
which, together with particle-hole symmetry, Eq. (\ref{particlehole})
implies $\sigma_y \boldsymbol{r}_h(\boldsymbol{k}_{||},\varepsilon)\sigma_y=\boldsymbol{r}_e(\boldsymbol{k}_{||},-\varepsilon)$ and thus the condition (\ref{isitgapped}) for a gapless system reduces to
\begin{equation}
\text{det}\Bigl(\boldsymbol{1}+\boldsymbol{r}_e^{-1}(\boldsymbol{k}_{||},0)\boldsymbol{r}_e(\boldsymbol{k}_{||},0)\Bigr)\neq 0, \quad \forall \boldsymbol{k}_{||},
\end{equation}
which is always fullfilled. Breaking of time-reversal symmetry is thus crucial in order to observe any topological phase transitions (which occur at gap-closings) in proximity-based superconducting 2D systems. Since no topological phase transitions can occur, we note that the system must always be in the trivial phase. We can thus conclude that, for proximity structures, the only \textit{accessible}Ê topological phase in symmetry classes CI and DIII is the trivial phase.

In the context of Majorana bound states it is furthermore crucial to also break spin-rotational symmetry since otherwise any quasiparticles in the system will have the form $\gamma=u\psi^\dag_\uparrow+v\psi_\downarrow$ or $\gamma=u\psi_\downarrow^\dag+v\psi_\uparrow$ neither of which can possess the defining property of Majorana fermions $\gamma^\dag=\gamma$. Thus it should be clear that the most interesting class (at least in the context of Majorana fermions) is the class D. Nevertheless, for pedagogical reasons we shall first consider the two classes CI and DIII in the next sections, followed by an analysis of the most interesting class D. While a complete analysis, including class C, would be desirable, it falls outside the scope of the present article. For interested readers we refer to the literature on proximity structures with orbital coupling to the magnetic fields \cite{VanOstaay2011,Zu2001,Giazotto2005,Rakyta2007}.

\subsection{Class CI}\label{CI}
\begin{figure}
\centering
\includegraphics[width=0.48\textwidth]{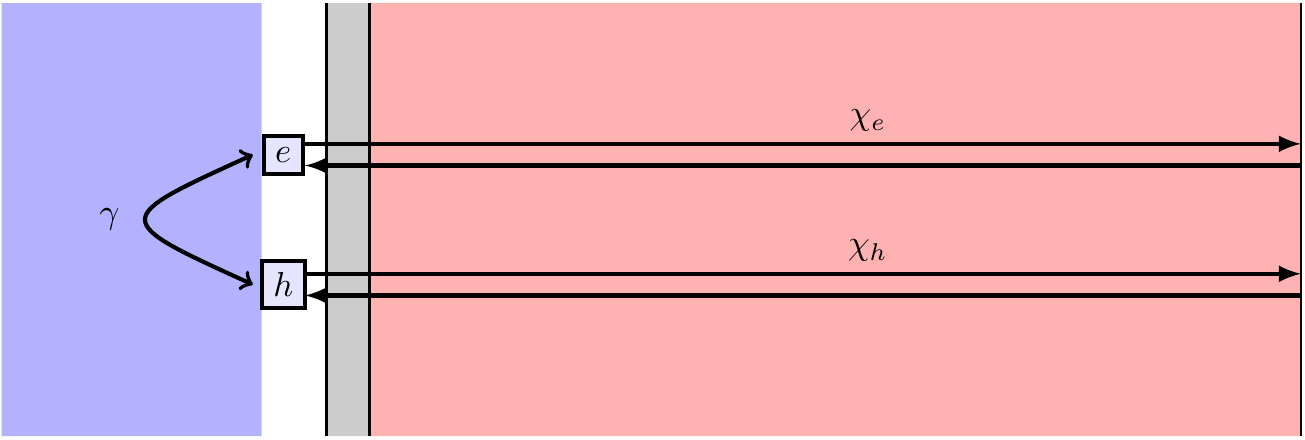}
\includegraphics[width=0.23\textwidth]{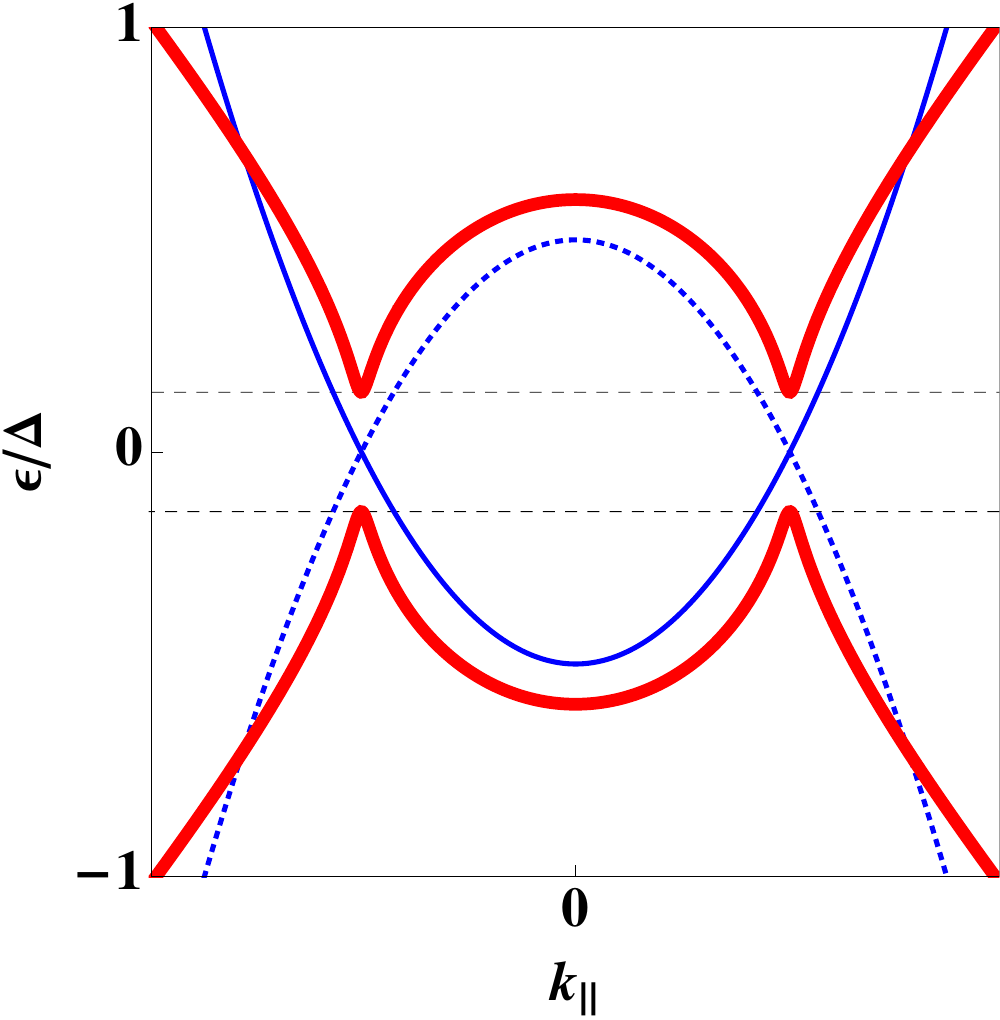}
\includegraphics[width=0.23\textwidth]{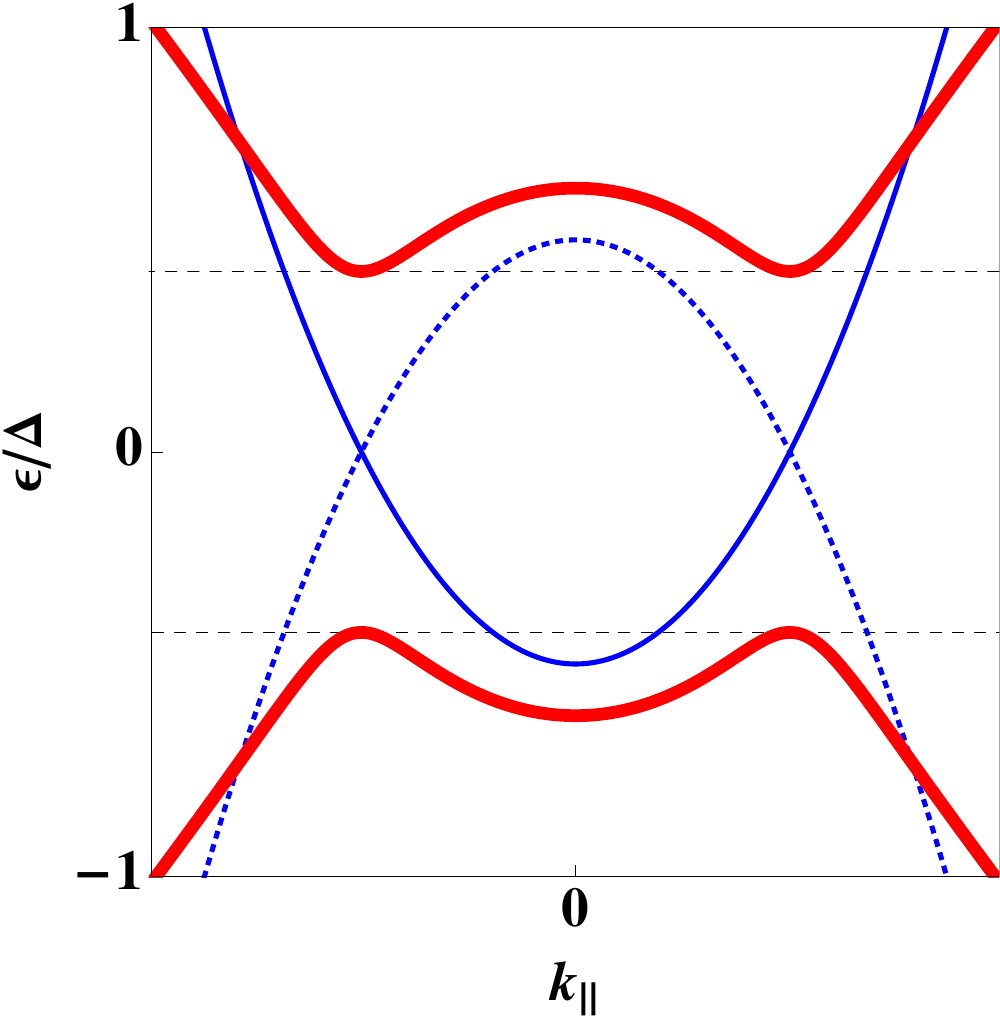}
\caption{(Top) Schematic illustration of the Andreev reflection process: Transmission through the semiconductor film and reflection at its surface produces an overall phase-shift $\chi_{e/h}$ for electrons/holes. Resonances in the film are related to these phase shifts through the Friedel sum rule (\ref{friedelsum}). Andreev reflection on the superconducting side connects the electron- and hole branches through an additional phase $\gamma$. A bound state is formed through the cyclic process of electron reflection on semiconductor film - Andreev reflection - hole reflection on semiconductor film - Andreev reflection - etc. The dispersion equation (\ref{andreev1}) can be interpreted as a type of Born quantization rule.
(Bottom:) Solution to the spectral equation (\ref{resonantsimple}) indicated by thick (red) line for weak (Left, $\Gamma=0.03\Delta$) and strong coupling (Right, $\Gamma=0.2\Delta$). The resonant bands are indicated by the parabolic thin and dashed curves (blue), while the effective gap $\Delta_\text{eff}$ is indicated by the horizontal dashed lines. Notice also the bending of the curves as the band approaches the bulk gap $\Delta$.}
\label{fig.resonantsimple}
\end{figure}

If the system preserves spin-rotational symmetry, the reflection matrix satisfies equation (\ref{spinrot}). Particle hole symmetry then requires $\boldsymbol{r}_h=e^{i\chi_h}\boldsymbol{1}$ with $\chi_h(\boldsymbol{k}_{||},\varepsilon)=\chi_e(-\boldsymbol{k}_{||},-\varepsilon)$. The spectral equation can then be written as 
\begin{equation}
(1-e^{-i(2\gamma(\varepsilon)+\chi_e-\chi_h)})=0.
\end{equation}
Time-reversal symmetry requires Eq. (\ref{timerev}), which implies $\chi_e(\boldsymbol{k}_{||},\varepsilon)=\chi_e(-\boldsymbol{k}_{||},\varepsilon)$, and $\chi_h(\boldsymbol{k}_{||},\varepsilon)=\chi_e(\boldsymbol{k}_{||},-\varepsilon)$. The spectral equation then reduces to the the "quantization condition"
\begin{equation}\label{andreev1}
2\gamma(\varepsilon)+\chi_e(\boldsymbol{k}_{||},\varepsilon)-\chi_e(\boldsymbol{k}_{||},-\varepsilon)=2\pi n.
\end{equation}
The physical interpretation of this equation can be understood from Fig. \ref{fig.resonantsimple} (Top).
To obtain a qualitative understanding of the solutions to this equation we may employ the Friedel sum rule:
\begin{equation}\label{friedelsum}
\chi_e(\boldsymbol{k}_{||},\varepsilon)-\chi_e(\boldsymbol{k}_{||},-\varepsilon)=\pi \int^{\varepsilon}_{-\varepsilon} d\varepsilon'\rho(\boldsymbol{k}_{||},\varepsilon'),
\end{equation}
where $\rho(\boldsymbol{k}_{||},\varepsilon)$ is the normal density of states in the film on top of the superconductor. If we assume that there exists a single (spin-degenerate) resonant band $\varepsilon_0(\boldsymbol{k}_{||})$ with a broadening $\Gamma$, i.e. the density of states has the generic form
\begin{equation}
\rho(\boldsymbol{k}_{||},\varepsilon)=\frac{1}{\pi}\frac{\Gamma}{\Gamma^2+(\varepsilon-\varepsilon_0(\boldsymbol{k}_{||}))^2},
\end{equation}
then the phase difference has the form
\begin{equation}\label{phasedifferenceresonant}\begin{split}
&\chi_e(\boldsymbol{k}_{||},\varepsilon)-\chi_e(\boldsymbol{k}_{||},-\varepsilon)=\\
&=\arctan\left(\frac{\varepsilon-\varepsilon_0(\boldsymbol{k}_{||})}{\Gamma}\right)+\arctan\left(\frac{\varepsilon+\varepsilon_0(\boldsymbol{k}_{||})}{\Gamma}\right).
\end{split}
\end{equation}
Taking the cosine of Eq. (\ref{andreev1}) and expanding for small $\Gamma$ one arrives at a BCS-like dispersion equation
\begin{equation}\label{resonantsimple}
\varepsilon^2=\varepsilon^2_0(\boldsymbol{k}_{||})+\Delta_{\text{eff}}^2, \qquad \Delta_{\text{eff}}=\sqrt{\Delta\Gamma}.
\end{equation}
The solution of the full dispersion equation is depicted in Fig. (\ref{fig.resonantsimple}) for the case of a resonance-band with parabolic dispersion $\varepsilon_0(\boldsymbol{k}_{||})=k_{||}^2/2m-\mu$. 
The main feature, of course, is the hybridization between the electron- and hole-branches of the resonant bands due to the Andreev reflection. The induced gap is given by $\Delta_{\text{eff}}=\sqrt{\Delta\Gamma}$. If there are multiple resonant bands in the film the problem can be treated analogously provided the energy-spacing between the bands is larger than the width $\Gamma$. The Andreev reflection will then induce hybridization between the electron- and hole branches of all bands (not necessarily at the chemical potential).

While this is a convenient and intuitive description of the problem, it does not provide any information about the bandstructure in terms of microscopic parameters. To this end, we model the semiconductor film region by an effective mass Hamiltonian and the interface between the superconductor and the film by a $\delta$-function potential of strength $V_0$. To account for the finite thickness $L$ of the film we impose hard-wall boundary conditions on the other boundary of the film. The scattering phase shift is then given by $\chi_e(\boldsymbol{k}_{||},\varepsilon)=\pi+2\delta_e(\boldsymbol{k}_{||},\varepsilon)$ with
\begin{equation}\label{phaseshift}
\tan\delta_e=\frac{v \sin(k'L)}{v'\cos(k'L)+2V_0\sin(k'L)},
\end{equation}
where $v(\boldsymbol{k}_{||},\varepsilon)=\text{Re}(k(\boldsymbol{k}_{||},\varepsilon))/m$, $v'(\boldsymbol{k}_{||},\varepsilon)=\text{Re}(k'(\boldsymbol{k}_{||},\varepsilon))/m'$ denote the velocities and $m$, $m'$ the masses in the superconductor and semiconductor regions respectively. The resonance condition $k'L=n\pi$ corresponds to resonance bands with energies $\varepsilon_{n,0}(\boldsymbol{k}_{||})=\tfrac{1}{2m'}\left(\tfrac{n\pi}{L}\right)^2+\tfrac{\boldsymbol{k}_{||}^2}{2m}-\mu$.
\begin{figure}
\includegraphics[width=0.23\textwidth]{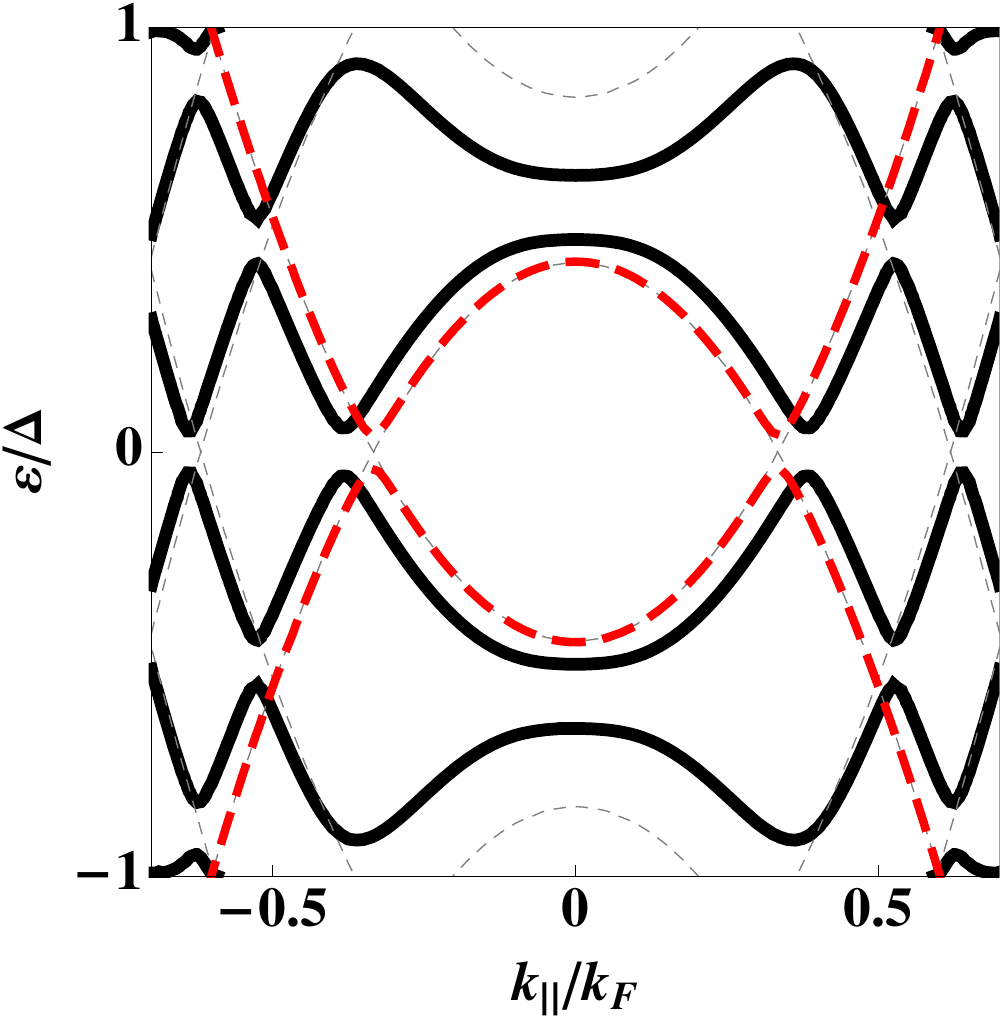}
\includegraphics[width=0.23\textwidth]{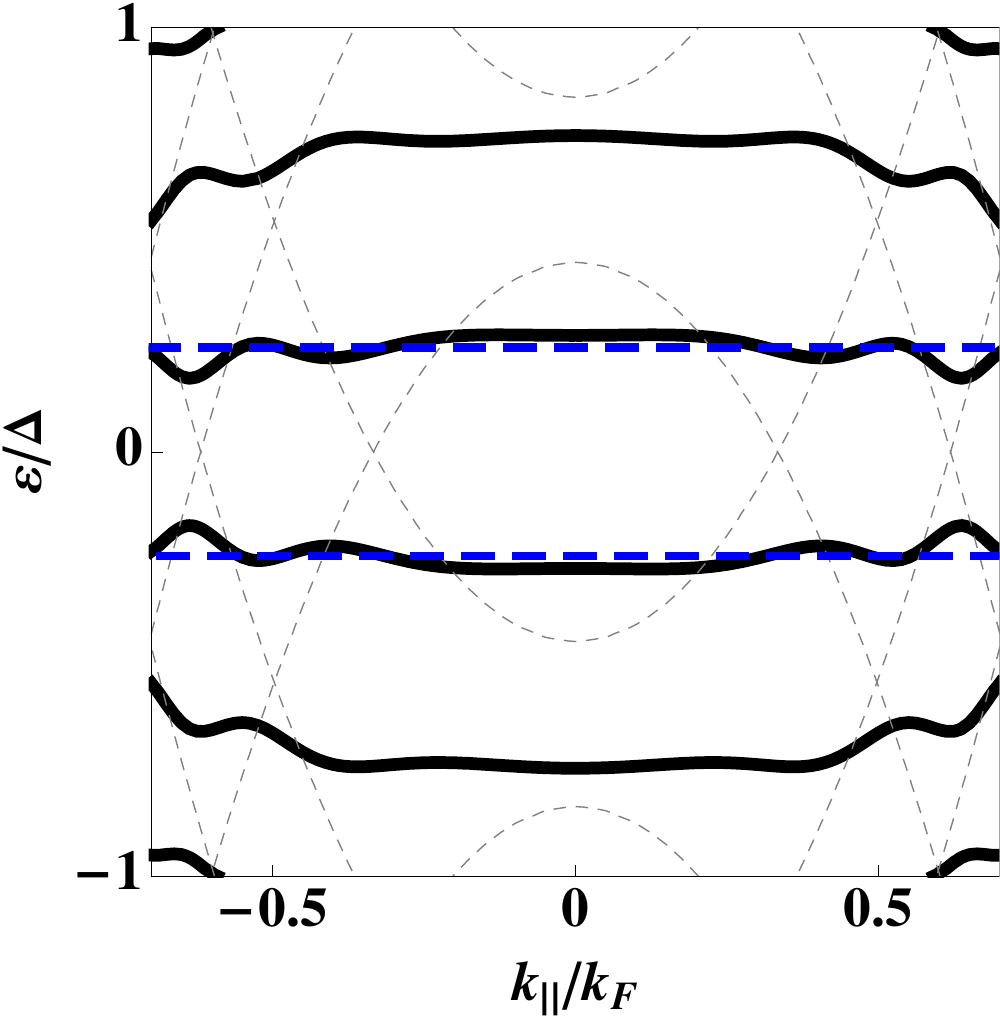}
\caption{2D Bandstructure of a typical film of thickness $L=3\lambda_F$ in class CI, i.e. respecting time-reversal and spin-rotational symmetry, are depicted with thick full lines, while the normal resonance-bands $\pm\varepsilon_{n,0}(\boldsymbol{k}_{||})$ are depicted by thin dashed lines. The thick dashed (red) lines show the dispersion from Eq. (\ref{inducedBCS}) with the approximated effective gap (\ref{effectivegap}). The fit works reasonably well for a weakly coupled system (left, $V_0=v_F$) while for a strongly coupled system (right, $V_0=0.05v_F$) the fit becomes progressively worse eventually approaching the solutions (\ref{deGsG}) obtained in the strong coupling limit of the spectral equation, here indicated in the right figure by the dashed (blue) horizontal lines.}
\label{fig.spectrumCI}
\end{figure}
Naively, one might try to expand the left hand side of Eq. (\ref{phaseshift}) around one of the resonances $\varepsilon_{n,0}(\boldsymbol{k}_{||})$ to obtain a form similar to Eq. (\ref{phasedifferenceresonant}), and thus obtain a spectral equation like Eq. (\ref{resonantsimple}). In this case the width of the resonance would be $\Gamma(\boldsymbol{k}_{||})=\tfrac{v'}{v}\tfrac{(\varepsilon(\boldsymbol{k}_{||})+\mu_{||})}{k'L}$. However, it turns out that the thusly evaluated effective gap, $\Delta_\text{eff}=\sqrt{\tfrac{v_F}{v_F'}\tfrac{\Delta\mu_{||}}{(k'_FL)}}$, neglects the effect of the interface potential $\propto V_0$, and thus overestimates the gap. By instead considering the low energy limit of the spectral equation, by keeping terms up to second order in $\varepsilon$, one obtains a much more realistic expression for the induced gap
which in the limit $2mV_0L\gg \text{max}[k_FL,\tfrac{\mu_{||}}{\Delta}\frac{k_FL}{2mV_0L}]$ reduces to
\begin{equation}\label{effectivegap}
\frac{\Delta_{\text{eff}}}{\Delta}=\frac{\sqrt{2}}{\sqrt{1+\tfrac{1}{2}\left(\frac{\Delta}{\mu_{||}}\right)^2(2mV_0 L)^2}}.
\end{equation}
With this expression for the effective gap, the expression
\begin{equation}\label{inducedBCS}
\varepsilon_{\pm}(\boldsymbol{k}_{||})=\pm \sqrt{\varepsilon^2_{0}(\boldsymbol{k}_{||})+\Delta_{\text{eff}}^2},
\end{equation}
fits reasonably well with the numerical solution for the lowest energy band around $\boldsymbol{k}_{||}=0$ (see Fig. \ref{fig.spectrumCI}) in the weak coupling limit $2V_0\gtrsim v_F$. In the opposite limit, with $v=v'$, we can expand the spectral equation in $\Delta/\mu$ (Andreev approximation) to recover the deGennes-Saint James states\cite{DeGennes1964} satisfying the spectral equation
\begin{equation}
\frac{\varepsilon}{\Delta}=\pm\cos\left(2\pi\tfrac{L}{\xi_0}\tfrac{\varepsilon}{\Delta}\right),
\end{equation}
with the approximate solution for the lowest energy states given by
\begin{equation}\label{deGsG}
\varepsilon=\pm\Delta\left(\frac{-1+\sqrt{1+2(2\pi \tfrac{L}{\xi_0})^2}}{(2\pi \tfrac{L}{\xi_0})^2}\right),
\end{equation}
where $\xi_0=\pi v_F/\Delta$ is the superconducting coherence length.
\subsection{Class DIII}
For a system which respects time-reversal symmetry, but not spin-rotational symmetry, the reflection matrix is constrained according to Eq. (\ref{timerev}) which together with particle hole symmetry reduces the spectral equation to
\begin{equation}\label{timerev3}
0=\text{det}\Bigl(\boldsymbol{1}-\boldsymbol{r}_e^{-1}(\boldsymbol{k}_{||},\varepsilon)e^{-i\gamma(\varepsilon)}\boldsymbol{r}_e(\boldsymbol{k}_{||},-\varepsilon)e^{-i\gamma(\varepsilon)}\Bigr).
\end{equation}
Let us assume, as is the case when the Hamiltonian in the semiconductor film contains a spin-orbit term, that the unitary matrix $\boldsymbol{U}$ which diagonalizes the reflection matrix $\boldsymbol{r}_e$ does not depend on energy:
\begin{equation}
\boldsymbol{r}_e(\boldsymbol{k}_{||},\varepsilon)=\boldsymbol{U}(\boldsymbol{k}_{||})e^{i\boldsymbol{\chi}(\boldsymbol{k}_{||},\varepsilon)}\boldsymbol{U}^\dag (\boldsymbol{k}_{||}),
\end{equation}
where $\boldsymbol{\chi}=\text{diag}(\chi_+,\chi_-)$ is a diagonal matrix containing the scattering phase shifts in the diagonalized basis, i.e. the corresponding phase-factors are the eigenvalues of the reflection matrix. If this is the case, then the same unitary matrix also diagonalizes the reflection matrix $\boldsymbol{r}_e(\boldsymbol{k}_{||},-\varepsilon)$, and the spectral equation reduces to
\begin{equation}
0=\text{det}\Bigl(\boldsymbol{1}-e^{-i\boldsymbol{\chi}(\boldsymbol{k}_{||},\varepsilon)}e^{-i\gamma(\varepsilon)}e^{i\boldsymbol{\chi}(\boldsymbol{k}_{||},-\varepsilon)}e^{-i\gamma(\varepsilon)}\Bigr).
\end{equation}
This equation is similar to the one in the class CI, with the important distinction that it decouples into two independent spectral equations
\begin{equation}\label{spectraleqDIII}
\begin{split}
&2\gamma(\varepsilon)+\chi_+(\boldsymbol{k}_{||},\varepsilon)-\chi_+(\boldsymbol{k}_{||},-\varepsilon)=2\pi n,\\
&2\gamma(\varepsilon)+\chi_-(\boldsymbol{k}_{||},\varepsilon)-\chi_-(\boldsymbol{k}_{||},-\varepsilon)=2\pi n.\\
\end{split}
\end{equation}
This situation applies to semiconductor films with spin-orbit interaction, or more generally to any Hamiltonian of the form $\boldsymbol{\mathcal{H}}(z>0)=(-i\partial_z)^2/2m'-\mu +\boldsymbol{b}(\boldsymbol{k}_{||})\cdot\boldsymbol{\sigma}$ with $\boldsymbol{b}(-\boldsymbol{k}_{||})=-\boldsymbol{b}(\boldsymbol{k}_{||})$. Such a Hamiltonian can be diagonalized by a unitary transformation which turns the vector $\boldsymbol{b}(\boldsymbol{k}_{||})$ into the direction of the spin quantization axis. This unitary matrix then also diagonalizes the reflection matrix, assuming the interface potential is spin-rotationally invariant. For each helicity quantum number $\sigma=\pm$ we have separate sets of normal resonance bands in the semiconductor film given by
\begin{equation}
\varepsilon_{n,\sigma,0}(\boldsymbol{k}_{||})=\frac{1}{2m}\left(\frac{n\pi}{L}\right)^2+\frac{\boldsymbol{k}^2_{||}}{2m}+\sigma|\boldsymbol{b}(\boldsymbol{k}_{||})|-\mu.
\end{equation}
For the specific case of Rashba spin-orbit interaction due to the structure inversion asymmetry in the $z$-direction we have $\boldsymbol{b}(\boldsymbol{k}_{||})=\alpha_{so}^R(k_y,-k_x,0)$, while for Dresselhaus spin-orbit interaction due to bulk inversion asymmetry in the same direction (the growth direction) we have $\boldsymbol{b}(\boldsymbol{k}_{||})=\alpha_\text{so}^D(k_x,k_y,0)$. 

The spectrum of the 2D semiconductor can now be evaluated by following the same line of arguments as in section \ref{CI}, with the important difference that there are now two independent branches of opposite helicity $\sigma=\pm$. Since the spectral equation (\ref{spectraleqDIII}) does not mix states of different helicity, only electron-hole branches within the same helicity-branch will hybridize. The absence of anti-crossing between electron-hole branches of opposite helicity is a direct consequence of the time reversal symmetry. Furthermore, two branches $\varepsilon_{n,+,0}(\boldsymbol{k}_{||})$ and $\varepsilon_{n,-,0}(\boldsymbol{k}_{||})$ cross at the time-reversal invariant momentum $\boldsymbol{k}_{||}=0$ as a consequence of $\boldsymbol{b}(-\boldsymbol{k}_{||})=-\boldsymbol{b}(\boldsymbol{k}_{||})$ as required by time-reversal invariance.
\begin{figure}
\includegraphics[width=0.23\textwidth]{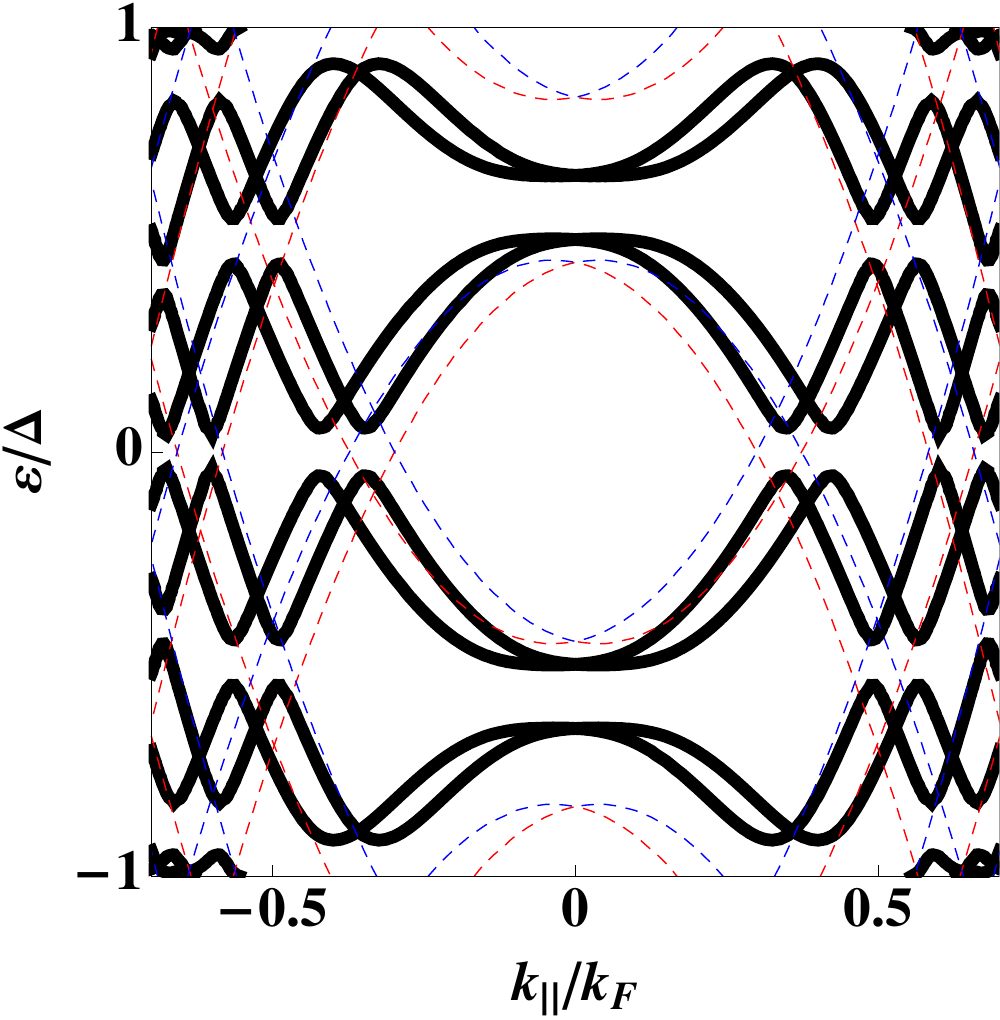}
\includegraphics[width=0.23\textwidth]{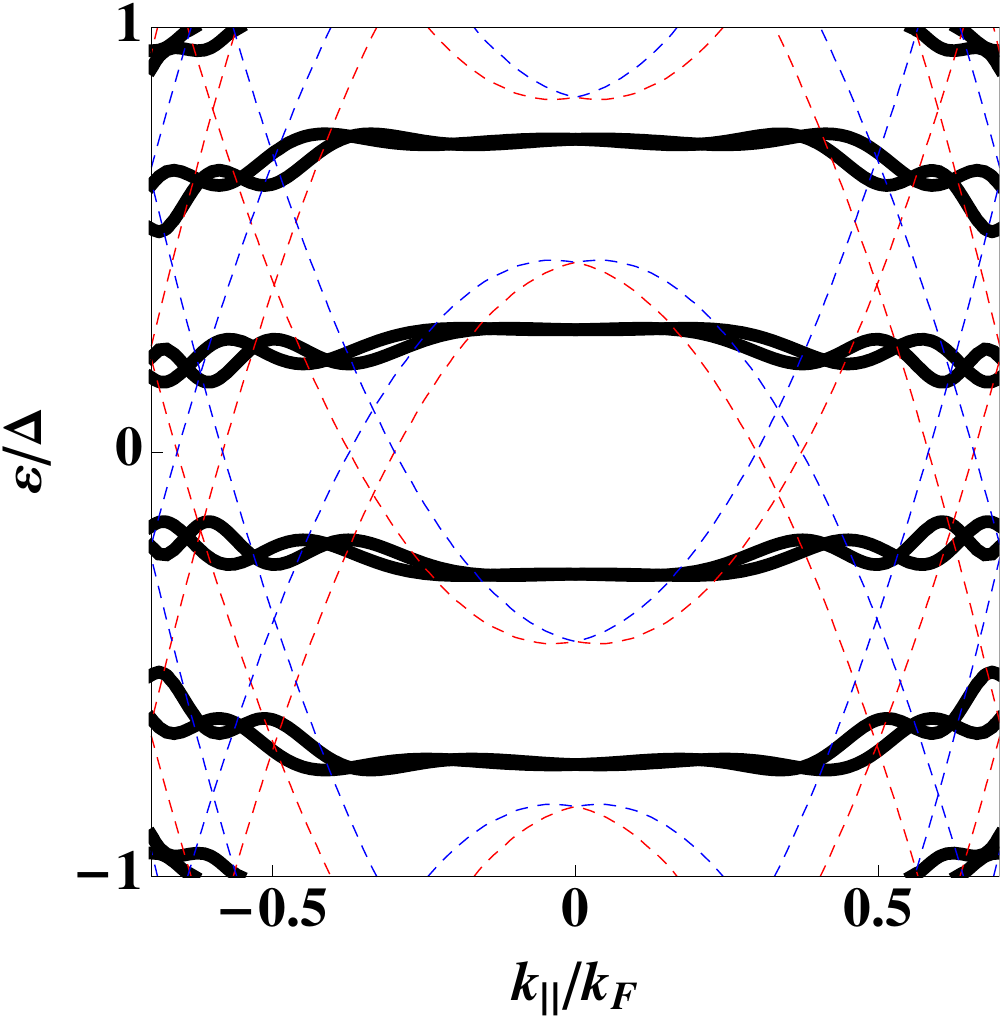}
\caption{2D Bandstructure (thick black lines) of a semiconductor film of thickness $L=3\lambda_F$ with Rashba spin-orbit interaction $\alpha_\text{so}^R=0.3\Delta/k_F$ in the weak (left, $V_0=v_F$) and strong (right, $V_0=0.1v_F$) coupling regime. The normal helical resonance-bands are depicted by thin dashed lines (red and blue).}
\label{fig.spectrumDIII}
\end{figure}
In the weak coupling regime the lowest energy bands can, for small $\boldsymbol{k}_{||}$ have the approximate form $\varepsilon_{\pm,\sigma}(\boldsymbol{k}_{||})=\pm \sqrt{\varepsilon^2_{n,\sigma,0}(\boldsymbol{k}_{||})+\Delta^2_\text{eff}})$ for the last normal state resonance band $n$ which crosses the Fermi-level and the effective gap given by Eq. (\ref{effectivegap}). This spectrum is consistent with a BCS model of the form $H=\sum_{\boldsymbol{k}_{||},\sigma}\left[\varepsilon_{\sigma,0}(\boldsymbol{k}_{||})c_{\sigma}^{\dag}(\boldsymbol{k}_{||})c_{\sigma}(\boldsymbol{k}_{||})\right. + \left. \Delta_{\text{eff}}c^\dag_{\sigma}(\boldsymbol{k}_{||})c^\dag_{\sigma}(-\boldsymbol{k}_{||})\right]$ and can thus be considered as induced equal-helicity pairing.  In the strong coupling regime the lowest energy bands again approach Eq. (\ref{deGsG}) (see Fig, \ref{fig.spectrumDIII}).
\subsection{Class D}
\begin{figure}
\includegraphics[width=0.23\textwidth]{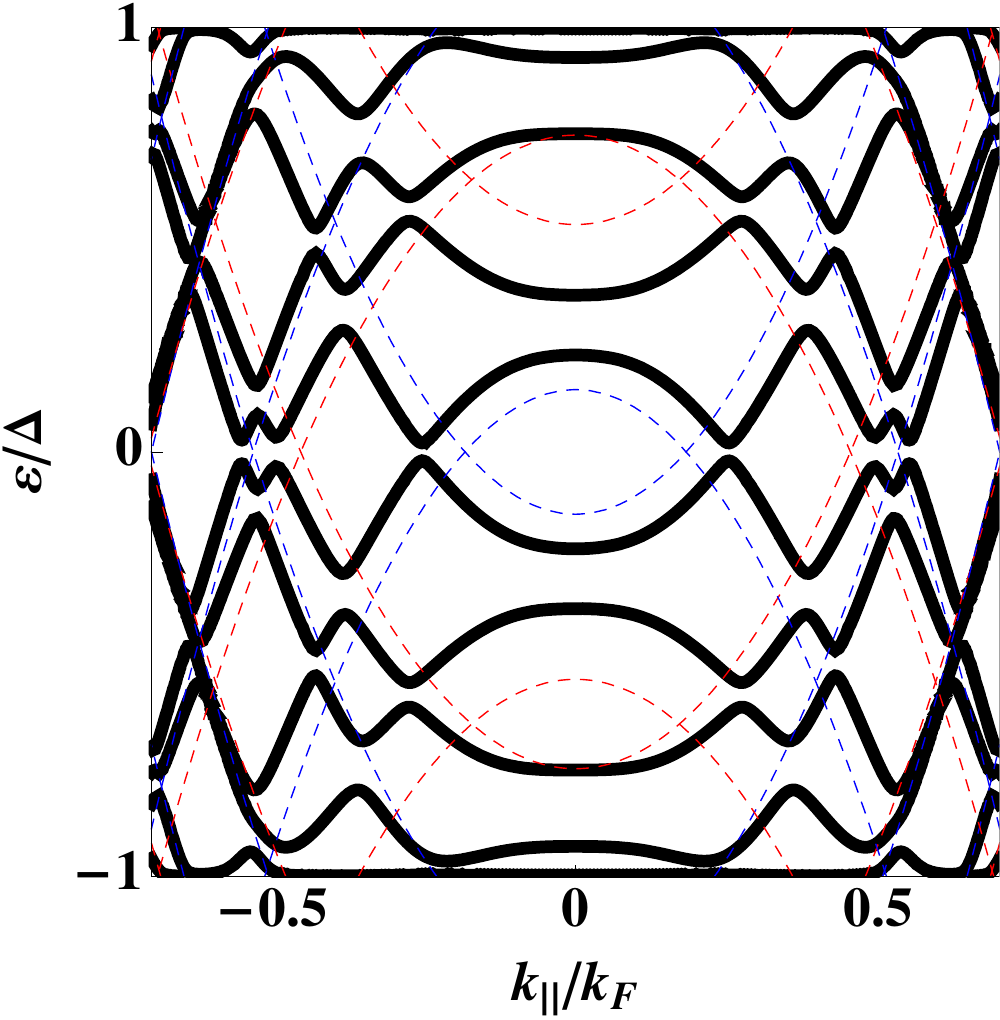}
\includegraphics[width=0.23\textwidth]{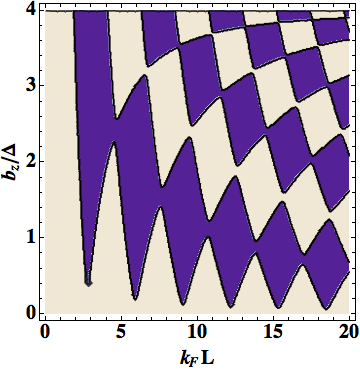}
\caption{(Left:) 2D Bandstructure (thick black lines) of a semiconductor film with Rashba spin-orbit interaction $\alpha_{\text{so}}^R=0.5\Delta/k_F$ and exchange/Zeeman splitting $b_z=0.3\Delta$ of thickness $L=3\lambda_F$ in the weak coupling regime. The normal helical resonance-bands are depicted by thin dashed lines (red and blue). (Right:) Gapped phases characterized by the $\mathbb{Z}_2$ invariant $\mathcal{M}$ from Eq. (\ref{topological2}) as a function of exchange splitting $b_z$ and thickness of the film $L$ in the weak coupling $(V_0=v_F$) regime. Here, beige refers to $\mathcal{M}=0$ while purple refers to $\mathcal{M}=1$.}
\label{fig.topological}
\end{figure}

Before proceeding to the main objective of this article - a semiconductor film with spin-orbit and exchange interaction - let us first consider the subclass of D where spin-rotational symmetry is not completely broken. This is the case if we have exchange interaction in the semiconductor film, and the normal state Hamiltonian in this region obtains a term $\propto \sigma_zb_z$, where $b_z$ denotes the strength of the exchange interaction. Since spin-rotational invariance is preserved around the $z$-axis the reflection matrix is diagonal and takes the form $\boldsymbol{r}_e(\boldsymbol{k}_{||},\varepsilon)=e^{i\boldsymbol{\chi}(\boldsymbol{k}_{||},\varepsilon)}$ with $\boldsymbol{\chi}=\text{diag}(\chi_\uparrow,\chi_\downarrow)$. Due to particle-hole symmetry we have $\boldsymbol{r}_h(\boldsymbol{k}_{||},\varepsilon)=e^{i\boldsymbol{\chi}(-\boldsymbol{k}_{||},-\varepsilon)}$, and the spectral equation becomes
\begin{equation}
0=\text{det}\Bigl(\boldsymbol{1}-e^{-i\boldsymbol{\chi}(\boldsymbol{k}_{||},\varepsilon)}\sigma_y e^{-i\gamma(\varepsilon)}e^{i\boldsymbol{\chi}(-\boldsymbol{k}_{||},-\varepsilon)}\sigma_ye^{-i\gamma(\varepsilon)}\Bigr),
\end{equation}
which again decomposes into two sets of spectral equations
\begin{equation}\label{spectraleqDlight}\begin{split}
&2\gamma(\varepsilon)+\chi_\uparrow (\boldsymbol{k}_{||},\varepsilon)-\chi_{\downarrow}(-\boldsymbol{k}_{||},-\varepsilon)=2n\pi,\\
&2\gamma(\varepsilon)+\chi_\downarrow (\boldsymbol{k}_{||},\varepsilon)-\chi_{\uparrow}(-\boldsymbol{k}_{||},-\varepsilon)=2n\pi.\\
\end{split}
\end{equation}
The effect of this spectral equation is that the electron-branch of the normal state resonance bands with spin $\uparrow$ will hybridize with the hole-branch of the normal state resonance bands with spin $\downarrow$ and vice versa. Spin-rotational invariance around the z-axis implies that spin is a good quantum number. Here we should consider that a hole with spin-down can be interpreted as a particle with spin-up and vice versa. This leads to the formation of bands with spin-up and spin-down without avoided level-crossings between the two. If we write the normal state resonance bands as $\varepsilon_{n,\sigma,0}(\boldsymbol{k})=\varepsilon_{n,0}(\boldsymbol{k}_{||})+\sigma b_z$, one finds that in the weak coupling regime the lowest energy bands have the approximate form
\begin{equation}
\varepsilon_{\pm,\sigma}(\boldsymbol{k}_{||})=\pm\sqrt{\varepsilon^2_{0}(\boldsymbol{k}_{||})+\Delta_\text{eff}^2}+\sigma b_z.
\end{equation}
From this equation it is clear that the main effect of the exchange interaction is to shift the energy levels up/down. At $|b_z|=\sqrt{\varepsilon^2_{0}(0)+\Delta^2_\text{eff}}\approx \mu_0$ we reach a point where two branches of opposite spin cross each other at $(\boldsymbol{k}_{||},\varepsilon)=(0,0)$. This feature can be investigated more carefully from the spectral equation (\ref{spectraleqDlight}) by setting $\varepsilon,\boldsymbol{k}_{||}=0$ for which we get $\chi_{\sigma}(0,0)-\chi_{-\sigma}(0,0)=(2n-1)\pi$. Applying again the Friedel sum rule we can write this condition as
\begin{equation}\label{topological}
N_{\uparrow}-N_{\downarrow}=(2n-1)\pi,
\end{equation}
where $N_\sigma$ is the number of normal state electronic states at $\boldsymbol{k}_{||}$ with spin $\sigma$ below $\varepsilon=0$. Thus the gap closing at this point coincides with the point where, in the normal state, the number of electronic spin-up bands minus the number of spin-down bands below the gap changes $\text{modulo }2$. This identification will be useful to us in the following when we combine spin-orbit interaction together with an exchange field. To this end, we write the normal state Hamiltonian in the semiconductor film as
\begin{equation}\begin{split}
&\boldsymbol{\mathcal{H}}=\tfrac{(-i\partial_z)^2}{2m'}-\mu_{||}+\boldsymbol{b}(\boldsymbol{k}_{||})\cdot\boldsymbol{\sigma},\\
&\boldsymbol{b}(\boldsymbol{k}_{||})=(\alpha_\text{so}^Rk_y,-\alpha_\text{so}^Rk_x,b_z).
\end{split}
\end{equation}
The last term clearly breaks spin-rotational symmetry, but also breaks time-reversal symmetry since $\boldsymbol{b}(-\boldsymbol{k}_{||})\neq -\boldsymbol{b}(\boldsymbol{k}_{||})$. The Hamiltonian can be diagonalized through a unitary matrix
\begin{equation}
\boldsymbol{U}=\begin{pmatrix}
\cos\frac{\theta}{2}e^{-i\varphi} & \sin\frac{\theta}{2}\\
-\sin\frac{\theta}{2} & \cos\frac{\theta}{2}e^{i\varphi}\end{pmatrix},
\end{equation}
where $(\sin\theta\cos\varphi,\sin\theta\sin\varphi,\cos\theta)=\boldsymbol{b}(\boldsymbol{k}_{||})/|\boldsymbol{b}(\boldsymbol{k}_{||})|$.
The same matrix also diagonalizes the reflection matrix $\boldsymbol{r}_e(\boldsymbol{k}_{||},\varepsilon)$ and thus we can write $\boldsymbol{r}_e=\boldsymbol{U}e^{i\boldsymbol{\chi}}\boldsymbol{U}^\dag$
where again $\boldsymbol{\chi}=\text{diag}(\chi_+,\chi_-)$ and $\chi_{\sigma}=\pi+2\delta_{\sigma}$ are defined analogously to (\ref{phaseshift}) but with $k'\rightarrow k'_\sigma$ for each helicity branch $\sigma=\pm$. Since time-reversal symmetry is broken, $\sigma_y\boldsymbol{r}_h\sigma_y$ can not be diagonalized by the same matrix $\boldsymbol{U}$. From the only remaining symmetry - particle-hole symmetry - we have
\begin{equation}
\boldsymbol{r}_h(\boldsymbol{k}_{||},\varepsilon)=\boldsymbol{U}^*(-\boldsymbol{k}_{||})e^{i\boldsymbol{\chi}(-\boldsymbol{k}_{||},-\varepsilon)}\boldsymbol{U}^T(-\boldsymbol{k}_{||}).
\end{equation}
\begin{figure}
\includegraphics[width=0.47\textwidth]{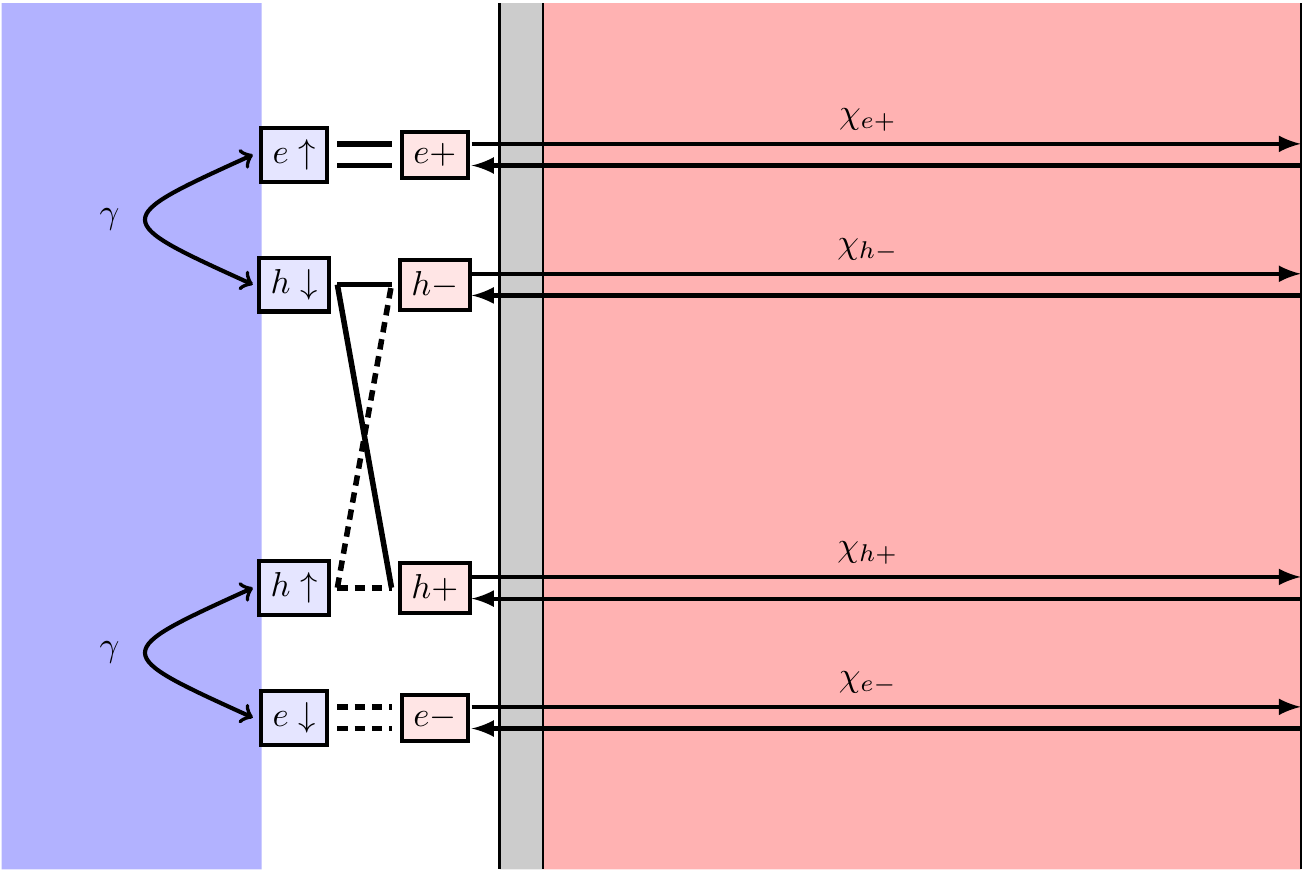}
\caption{Andreev reflection processes in the presence of spin-orbit interaction and Zeeman/exchange splitting. The normal state resonant bands are associated with the phase-shifts corresponding to the eigenvalues of the reflection matrices $\boldsymbol{r}_e$ and $\boldsymbol{r}_h$. Due to the absence of both spin-rotational symmetry and time-reversal symmetry, diagonalizing $\boldsymbol{r}_e$, here indicated by straight lines in the intermediate (white) region, does not simultaneously diagonalize $\boldsymbol{r}_h$, thus depicted in the figure by crossed- and straight lines. Thus closed cycles involve electron- and hole branches, which become mixed by Andreev reflection, of both helicities. The mixing is determined by the angle $\theta$ in Eq. (\ref{spectrumfinal}).  Consequently there is a hybridization of electron-hole branches between all resonant states. The exception happens at $\boldsymbol{k}_{||},\varepsilon=0$ where spin-orbit interaction is absent. }
\label{andreevcomplex}
\end{figure}
Using this parametrization and evaluating the determinant directly one obtains the following spectral equation
\begin{equation}\label{spectrumfinal}\begin{split}
 0&=\cos^2\theta \sin(\gamma-\delta\chi^s_{+})\sin(\gamma-\delta\chi^s_{-})\\
 &+\sin^2\theta \sin(\gamma-\delta\chi^t_{+})\sin(\gamma-\delta\chi^t_{-}),
 \end{split}\end{equation}
where we have defined
\begin{equation}\begin{split}
\delta\chi^s_{\sigma}(\boldsymbol{k}_{||},\varepsilon)&=\chi_{\sigma}(\boldsymbol{k}_{||},\varepsilon)-\chi_{-\sigma}(-\boldsymbol{k}_{||},-\varepsilon),\\
\delta\chi^t_{\sigma}(\boldsymbol{k}_{||},\varepsilon)&=\chi_{\sigma}(\boldsymbol{k}_{||},\varepsilon)-\chi_{\sigma}(-\boldsymbol{k}_{||},-\varepsilon).\\
\end{split}
\end{equation}
The subscripts $s$ and $t$ indicate hybridization between electron-hole branches of opposite and equal helicities and thus, in a sense, induced singlet-, or  triplet superconductivity respectively. Setting $\varepsilon=0$ in this spectral equation we can identify points where the gap closes. Using $\gamma(\varepsilon=0)=\pi/2$ along with the fact that the scattering phase shifts only depend on $|\boldsymbol{b}(\boldsymbol{k}_{||})|$ and everything else is isotropic, i.e. $\chi_{\sigma}(-\boldsymbol{k}_{||},\varepsilon)=\chi_{\sigma}(\boldsymbol{k}_{||},\varepsilon)$ and thus $\delta\chi^t_{\sigma}(\varepsilon=0)=0$ and $\delta\chi^s_\sigma(\varepsilon=0)=-\delta\chi^s_{-\sigma}(\varepsilon=0)$, we get
\begin{equation}\begin{split}
 0&=\cos^2\theta(\boldsymbol{k}_{||}) \cos^2\Bigl(\chi_{+}(\boldsymbol{k}_{||},0)-\chi_-(\boldsymbol{k}_{||},0)\Bigr) +\sin^2\theta(\boldsymbol{k}_{||}).
 \end{split}\end{equation}
Since both terms on the right hand side are positive, the spectral equation can only be satisfied if both are equal to zero at the same time. This then reduces to the two constraints which have to be simultanously fulfilled:
\begin{equation}
\sin\theta(\boldsymbol{k}_{||})=0, \qquad \cos\Bigl(\chi_{+}(\boldsymbol{k}_{||},0)-\chi_-(\boldsymbol{k}_{||},0)\Bigr).
\end{equation}
The first one is only zero at $\boldsymbol{k}_{||}=0$ we find that this is the only place where a gap-closing may occur. Furthermore, the second condition is the same condition as discussed in the beginning of this section and is thus determined by Eq. (\ref{topological}), i.e. the gap closes whenever the parity of the difference of occupied normal state spin-up/spin-down states is changed. Thus the gap closing is associated with a band-inversion and the different gapped phases can be associated with the integer number $n\in \mathbb{Z}$ from Eq. (\ref{topological}). Distinguishing in particular odd, and even parity we can define a $\mathbb{Z}_2$ number
\begin{equation}\label{topological2}\begin{split}
\mathcal{M}&=(N_+-N_-)\text{mod } 2\\
&=\text{sign}\Bigl(\cos\bigl(\chi_+(0,0)-\chi_-(0,0)\bigr)\Bigr)\in \mathbb{Z}_2.
\end{split}
\end{equation}
This number, although not derived directly by topological arguments, can serve as a topological invariant in the sense that the number does not change within a gapped phase, only when the gap is closed can we go from odd or even parity. The corresponding bandstructure and "topological phase-diagram" is depicted in Figure \ref{fig.topological}. Interestingly this number does not depend on the properties of the bulk superconductor but only on the properties of the semiconductor film (of course a coupling to the superconductor must exist or else the formalism becomes meaningless).
\begin{figure}
\includegraphics[width=0.23\textwidth]{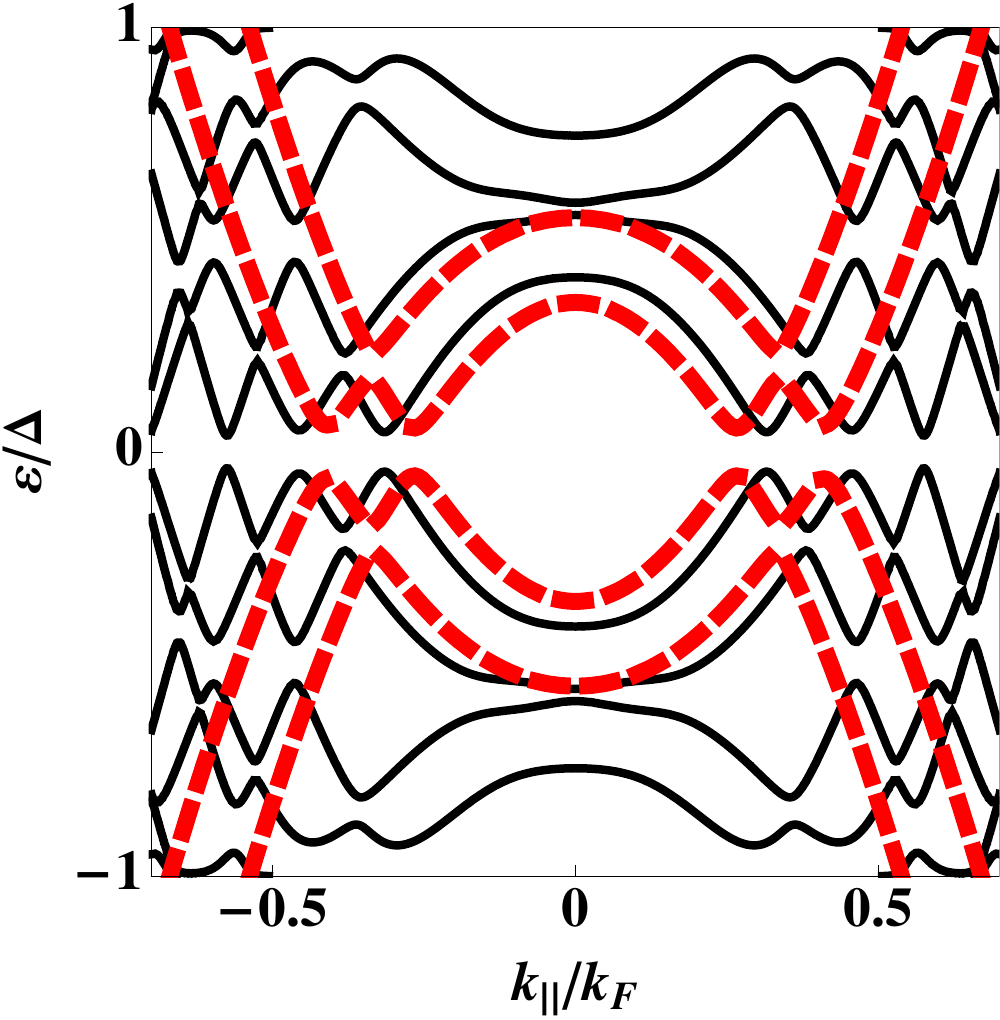}
\includegraphics[width=0.23\textwidth]{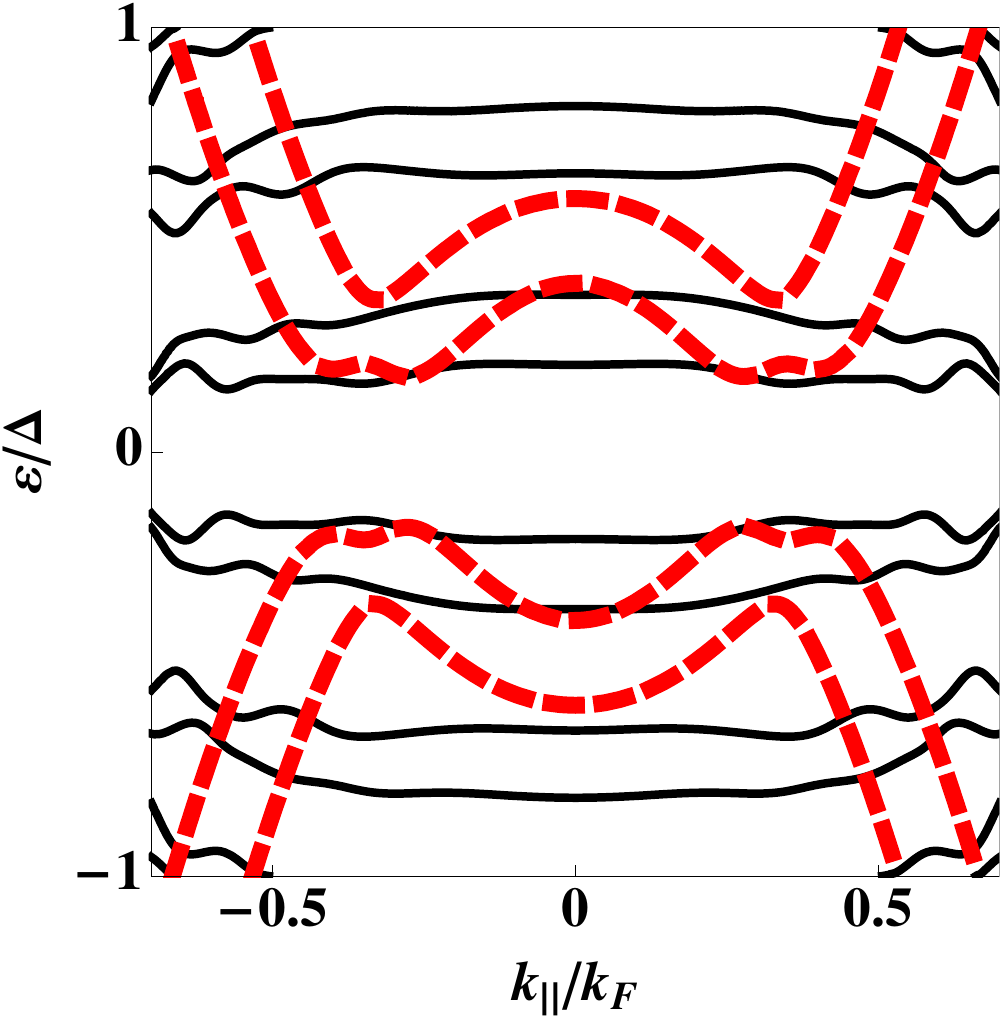}
\caption{2D Bandstructure (black lines) of a semiconductor film with Rashba spin-orbit interaction $\alpha_{\text{so}}^R=0.5\Delta/k_F$ and exchange/Zeeman splitting $b_z=0.1\Delta$ of thickness $L=3\lambda_F$ in the weak (left, $V_0=v_F$) and strong (right, $V_0=0.1v_F$) coupling regime and the dispersion obtained from the effective model Eq. (\ref{mappingtoSOmodel}) depicted by dashed red lines.}
\label{fig.mappingtoSOmodel}
\end{figure}
For weak coupling, the four lowest energy bands approximately follow the dispersion relation
\begin{equation}\label{mappingtoSOmodel}\begin{split}
\varepsilon^2_{\pm}=&b_z^2+\Delta^2_\text{eff}+\xi^2_{\boldsymbol{k}_{||}}+|\alpha_\text{so}k_{||}|^2\\
&\pm 2\sqrt{b^2_z\Delta_\text{eff}^2+b_z^2\xi_{\boldsymbol{k}_{||}}^2+|\alpha_{\text{so}}k_{||}|^2\xi_{\boldsymbol{k}_{||}}^2},
\end{split}
\end{equation}
derived\cite{Oreg2010} from the phenomenological model where a finite pair potential is introduced by hand, with the induced pair potential given by the approximate expression (\ref{effectivegap}). The fit works well for weak coupling, but breaks down for stronger coupling (see Fig. \ref{fig.mappingtoSOmodel}).
\section{Conclusions}
We have presented a general method of extracting the 2D bandstructure of films deposited on top of a conventional superconductor. Using the symmetry classification of Altland and Zirnbauer we concluded, based on general arguments, that for classes CI (spin-rotational symmetry and time-reversal symmetry) and C (spin-rotational symmetry only) the only accessible topological phase is the trivial one, while in class D there exists a $\mathbb{Z}_2$ number which can be expressed through the difference of spin-up vs spin-down scattering phase shifts at the particle-hole invariant point $\boldsymbol{k}_{||}=0=\varepsilon$ as $\mathcal{M}=\text{sign}(\cos(\chi_+(0,0)-\chi_-(0,0)))$. The change in this number is accompanied by a closing of the gap in the 2D bandstructure, in accordance with the typical notion of a topological phase transition. We furthermore derived analytical and numerical results for the 2D bandstructure in materials with spin-orbit coupling and/or Zeeman/exchange-splitting. 

\bibliographystyle{h-physrev}	
\bibliography{library}		

\end{document}